# CHARGING OF INTERSTELLAR DUST GRAINS NEAR THE HELIOPAUSE

Qianyu Ma, Lorin Matthews, Victor Land, Truell Hyde
Center for Astrophysics, Space Physics and Engineering Research,
Baylor University, Waco, TX, 76798-7316 USA
E-mail: qianyu_ma@baylor.edu

## ABSTRACT

The deflection of interstellar dust grains in the magnetic field near the heliopause has been investigated based on the assumption that interstellar grains are homogeneous spheres. However, remote observations have shown that interstellar grains are more likely to be composites of a large number of subunits. This has profound significance when interpreting data obtained through *in-situ* measurements, for the deflection of interstellar grains depends on their charge-to-mass ratio, and aggregates acquire different surface charges from spheres due to their complex structure. In this paper, the charging of aggregates near the heliopause is examined including both plasma charging and secondary electron emission. The results show that aggregates generally have a higher charge-to-mass ratio than spheres, and the small particle effect from secondary electron emission is evident for aggregates consisting of nano-sized particles. A new approach to estimate the aggregate charge with the aid of its structural characteristics is presented. The charge-to-mass ratio is used to derive the mass distribution of interstellar dust near the heliopause, and the result shows an overall agreement with Ulysses data.

## 1. INTRODUCTION

Dust is a ubiquitous component of the universe. It plays an important role in the thermodynamics and chemistry of the interstellar medium, in interstellar gas dynamics, and in the formation of stars, planets, and planetesimals. Progress in understanding the nature of interstellar dust not only provides important information on a significant constituent of the universe, but also helps astrophysicists to better understand the interstellar medium and the formation of stars and planets.

Interstellar dust has been studied through both remote observations and *in-situ* measurements. The Ulysses mission was the first mission which provided reliable detection of interstellar dust (Grün et al. 1993; Krüger et al. 2001). However, a problem arose when comparing Ulysses data to those of remote observations. The detected interstellar dust mass $m$ ranged from $10^{-20}$ to $10^{-11}$ kg with a maximum around $10^{-17} – 10^{-16}$ kg (Kimura et al. 2003). In contrast, the mass range for interstellar dust derived from the interstellar extinction and polarization observations indicates most particles being smaller than $10^{-16}$ kg (Mathis et al. 1977; Draine & Lee 1984; Désert et al. 1990). Since there is no clear evidence showing that the size distribution of the local interstellar medium close to the solar system is different from that of the average interstellar medium, this discrepancy has been attributed to the filtration process as the interstellar dust crosses the heliopause to enter the heliosphere.

As grains enter the solar system from beyond the heliopause, they are subject to the solar radiation pressure and solar gravitational force. The ratio of these two forces, $\beta$, is a measure of this relative importance for dynamics of small grains. For interstellar dust grains with $m < 10^{-17}$ kg, $\beta$ is less than unity (Kimura et al. 2003), implying that the depletion of small interstellar grains in the heliosphere is not due to the radiation pressure.

As grains enter the heliopause, they are also influenced by the force of the magnetic field, which provides a possible mechanism for the depletion of small interstellar dust particles (Levy & Jokipii 1976; Kimura & Mann 1998; Kimura & Mann 2000; Linde & Gombosi 2000). Between the heliopause and termination shock, the interaction of the solar wind and interstellar medium causes a sharp increase in the plasma temperature (Pauls & Zank 1996). The enhanced secondary electron emission is the dominant charging process for interstellar dust crossing the heliopause (Kimura & Mann 1998). Charged interstellar dust grains are deflected from their initial paths under the influence of the Lorentz force, with sufficiently large deflections preventing interstellar dust grains from reaching the inner solar system, and being detected by a spacecraft such as Galileo or Ulysses. Recent simulations of the dynamics of dust grains in the interstellar medium and the heliosphere have demonstrated that dust grains do experience a filtration process in the region of the heliopause (Linde & Gombosi 2000). While these results are consistent with satellite observations, the predicted cutoff-mass is approximately one order of magnitude lower than that observed by the in-situ measurements made by Ulysses.

This discrepancy may be due to the fact that dust particles with small mass experience a stronger Lorentz force than previously suggested, which implies a higher charge-to-mass ratio for these grains. Most of the previous work on the charging of interstellar dust has been based on the assumption that the interstellar grains are spherical in shape. However remote measurements have shown that the interstellar grains are more likely to be fluffy aggregates consisting of many tiny particles (Mathis & Whiffen 1989; Woo et al 1994). The grain sizes range from 5 nm to 0.25 μm (Mathis et al. 1977), with the upper limit undetermined, since large grains make only minor contributions to the interstellar extinction curve.

In this paper, a 3D model is employed to calculate the charge on aggregates near the heliopause. This is the first study to examine the charge on aggregates taking into account structural characteristics. The charge is determined by including both the plasma current and secondary electron emission. We demonstrate that aggregates consisting of nano-sized grains can acquire a significantly higher charge-to-mass ratio, as compared to spheres with the same mass. This is the result of the small particle effect from secondary electron emission, in which the grain charge is enhanced for the monomer radius $a < 10$ nm (Chow et al. 1993). The mass distribution of interstellar dust near the heliopause based on the aggregates' charge-to-mass ratio is compared to the previous model. The current model shows an overall agreement with the Ulysses data, although it does deviate for both small mass and large mass particles.

## 2. CHARGING MODEL AND PARAMETERS

The charge on a dust grain embedded in plasma is determined by

$$\frac{dQ}{dt} = \sum_j I_j, \tag{1}$$

where $I_j$ is the current contributed by the $j^{th}$ charging process. The charge on the dust grain reaches equilibrium when $\sum I_j = 0$. Here we consider two main charging processes: (I) collection of plasma particles, and (II) secondary electron emission.

Although interstellar dust grains at the heliopause are also exposed to the solar flux and interstellar radiation, photoemission is negligible compared to secondary electron emission (Kimura & Mann 1998), and thus neglected in this study.

The composition of interstellar dust grains is still unclear. However, silicates have been widely accepted as a major constituent of interstellar dust (Savage & Mathis 1979; McCarthy et al. 1980). Silicate monomers with radii from 5 nm to 500 nm were used in the simulation, with a density of 3.2 g cm$^{-3}$ (Draine & Salpeter 1979). The plasma density and temperature near the heliopause are set to be $2 \times 10^5$ m$^{-3}$ and $2 \times 10^6$ K (Pauls & Zank 1996). Only electrons and singly ionized hydrogen are considered. Other plasma components are neglected due to their relatively small contribution (Schwenn 1990).

## 2.1. Collection of Plasma Particles

The current to a dust grain can be found from Orbital Motion Limited theory (OML), which is based on the conservation of energy and angular momentum. The current density to any point on the surface of a dust grain due to the collection of a given species of plasma particles is defined as (Whipple 1981)

$$J_s = n_s q_s \iiint v_s f(\vec{v}_s) \cos\alpha \, d\vec{v}_s^{\,3}, \qquad (2)$$

where $n_s$ and $q_s$ are the number density and charge of the given species, $v_s$ is the speed of the particles, $f(v_s)$ is the distribution function which is assumed to be Maxwellian (Goertz 1989), and $\alpha$ is the angle between the impinging velocity and the surface normal of the dust grain. In the three dimensional case, we use spherical coordinates

($v$, $\theta$, $\phi$) in $\vec{v}$ space (Laframboise & Parker 1973). This allows the integration over the speed to be separated from the integral over the angles, and the differential velocity $d\vec{v}_s^3$ can be written as

$$d\vec{v}_s^3 = v^2 dv d\Omega, \quad (3)$$

which allows Eq. (2) to be rewritten as

$$J_s = n_s q_s \int v_s^3 f(v_s) dv_s \iint \cos\alpha \, d\Omega. \quad (4)$$

While the integration over speed is easy to carry out, the differential solid angle $d\Omega$ requires numerical simulation for aggregates, which is discussed in Section 2.3.

**2.2. Secondary Electron Emission**

Energetic primary electrons can release secondary electrons from the surface of a dust grain upon impact, which constitutes a positive charging current. It has been shown that the secondary electron yield is enhanced when the dimensions of the grains are comparable to the primary electron penetration depth, the so-called small particle effect (Chow et al. 1993). Since the size of a representative interstellar dust grain is normally less than 10 μm, we employ a model which takes the small-particle effect into account in determining the yield as a function of $E_0$, the initial energy of the incident particle (Draine & Salpeter 1979),

$$\delta(E_0) = \delta_m \frac{8(E_0/E_m)}{(1+E_0/E_m)^2}[1-\exp(\frac{-4a}{3\lambda})]f_1(\frac{4a}{3R})f_2(\frac{a}{\lambda}). \quad (5)$$

Here

$$f_1(x) = \frac{1.6 + 1.4x^2 + 0.54x^4}{1 + 0.54x^4}$$
$$f_2(x) = \frac{1 + 2x^2 + x^4}{1 + x^4},$$
(6)

and $a$ is the size of the grain. The maximum yield $\delta_m$, and the corresponding maximum energy $E_m$, are 2.4 and 400 eV for silicate (Mukai 1981). The escape length $\lambda$ is 2.3 nm (Draine & Salpeter 1979). The projected range $R$ gives the penetration of a primary electron into matter along the incident direction, and is determined based on $E_0$ as shown by Draine & Salpeter (1979).

Thus, the current due to secondary electron emission is calculated as

$$J_{sec} = n_e q_e \iiint v \cos\alpha f(v) \delta(E_0) d\vec{v}^3 \times \int_{E_{min}}^{\infty} \rho(E) dE, \quad (7)$$

where $\rho(E)$ is the energy distribution of the emitted electrons. It can be written as

$$\rho(E) = \frac{E}{2(kT_{sec})^2} [1 + \frac{1}{2}(\frac{E}{kT_{sec}})^2]^{-3/2}, \quad (8)$$

where $T_{sec}$ is the temperature of the emitted electrons and is set to be 2 eV (Goertz 1989). The lower limit of the integral is $E_{min} = \max(0, e\varphi)$, with $\varphi$ being the surface potential of the target grain. Eqs. (3) and (7) can be combined to yield

$$J_{sec} = n_e q_e \int v^3 f(v) \delta(E_0) dv \iint \cos\alpha d\Omega \int_{E_{min}}^{\infty} \rho(E) dE, \quad (9)$$

which has a form similar to that of Eq. (4).

## 2.3. Line-of-sight Approximation

Although the equilibrium charge on a single sphere embedded in plasma can be determined analytically by substituting Eqs. (4) and (9) into Eq. (1), the equilibrium charge on an aggregate can only be obtained through numerical simulation given its complex structure. The key to calculating the charge on an aggregate is determining the solid angle $d\Omega$ in Eqs. (4) and (9). The charging code *OML_LOS* calculates the electron and ion fluxes by determining the open lines of sight (LOS) to many points on the surface of each constituent monomer (Matthews & Hyde 2008). Incoming electrons and ions are assumed to move in a straight line and are captured at the points at which they intersect a monomer if their paths are not blocked by any monomers, including the target monomer. The charge on an aggregate is approximated using a multipole expansion. In the current study, only the monopole and dipole contributions are considered (Matthews & Hyde 2008; Matthews & Hyde 2009).

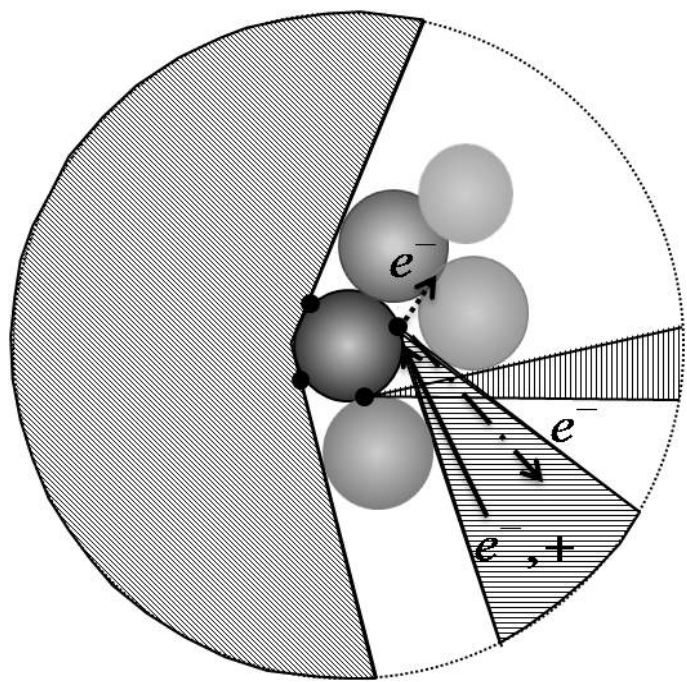

Fig. 1. Open lines of sight to given points on a monomer in an aggregate are indicated by the shaded regions. Charging currents to a given point are only incident from these directions. The dotted line indicates an emitted electron which is recaptured by another monomer along a closed line of sight, while the dash dotted line indicates an emitted electron that escapes along a free line of sight.

As shown in Fig. 1, the open lines of sight for specific given points on the surface of a monomer within an aggregate are used to approximate the solid angle in the integral in Eqs. (4) and (9). The surface of each monomer is divided into equal-area patches. Test directions from the center of each patch (the so called lines of sight), are determined to be *blocked* if they intersect any other monomer in the aggregate, or the monomer in question ($LOS_t = 0$), and *open* otherwise ($LOS_t = 1$). The line-of-sight factor is equal to the sum of the open lines of sight multiplied by the cosine of the angle of the test direction with the normal, and by the area of the patch on a unit sphere, $LOS = \iint \cos\alpha \, d\Omega = \sum_t LOS_t \cos\alpha_t \Delta(\cos\theta)\Delta\phi$. The factor LOS then replaces the integration over the angles in calculating the current density, given by Eqs. (4) and (9).

The net current of species $s$ to a given patch at a given time, $I_s(t)$, is then found by multiplying the current density by the area of the patch, $A$: $I_s(t)=J_s(t)A$. Summing over the species $s$ provides the change in the surface charge on the patch during a time interval $dt$, $dQ(t)=\Sigma I_s(t)dt$. The contribution to the dipole moment is $d\mathbf{D}(t)=\Sigma I s(t)\mathbf{R}dt$, where $\mathbf{R}$ is the distance vector from the patch to the center of the grain. Note that the current density depends on the potential of the grain, which in turn depends on the charge and dipole moment on each monomer, so that the solution requires numerical iteration until equilibrium is reached. The change in the charge and dipole moment of each monomer is then obtained by adding up the contribution of all the patches. The change in the charge and dipole of the aggregate is obtained by adding the contribution from each of the $N$ monomers. This process is iterated in time until the change in aggregate charge becomes negligible, $dQ_{agg} < 0.001Q_{agg}$, at which point on average the net current to the aggregate will be near zero.

Similarly, when secondary electrons are released from the surface of a monomer, a random test direction is chosen. The electron escapes the surface of the aggregate only if that direction is along an open line of sight. Electrons which are released along a blocked line of sight are recaptured by another monomer within the aggregate, leaving the total charge of the aggregate unchanged, but the charge distribution on the surface is altered.

## 2.4. Aggregate Builder and Compactness Factor

The numerical code *Aggregate Builder* is used to create aggregates through the coagulation of mono-disperse spheres using a combination of particle-cluster aggregation (PCA), and cluster-cluster aggregation (CCA) (Matthews et al. 2007; Matthews & Hyde 2009). During PCA, a target particle is placed at the origin, and a single particle is released at the boundary of the simulation box with its speed determined by Brownian motion and directed towards the center plus an offset. The trajectory of the incoming aggregate is calculated based on the electrostatic force acting on the aggregate, while the orientations of the particles are determined from the torques due to the charge-dipole interaction (Matthews et al. 2007; Matthews & Hyde 2009). A successful collision is detected if constituents of the target and projectile actually touch or overlap. The grains are assumed to have relative velocities that are too low for any restructuring to occur, and to stick at the point of contact (Wurm & Blum 1998; Blum & Wurm 2000). New aggregate parameters are then calculated, including the new charge as determined by OML_LOS. The resultant aggregate is saved to a library, with target aggregates allowed to grow until the number of monomers reaches twenty. In the case of CCA, small aggregates from the previously saved library are employed as the target grains, with the incoming grain either a spherical monomer or an aggregate randomly selected from the same library. The simulation continues as above until the number of monomers reaches 200. The radii of the monomers used for this simulation are 5 nm, 10 nm, 50 nm, 100 nm and 500 nm.

The fluffiness of aggregates is an essential parameter as more open-structured aggregates will have greater surface area, thus possessing more charge. Here we use the compactness factor $\Phi_\sigma$ to describe the fluffiness of the aggregate (Paszun & Dominik 2006),

$$\Phi_\sigma = N\left(\frac{a}{R_\sigma}\right)^3, \tag{10}$$

where $N$ is the number of monomers in the aggregate, $a$ is the monomer radius, and $R_\sigma$ is the radius of the averaged projected surface area, defined as

$$R_\sigma = \sqrt{\frac{\sigma}{\pi}}, \tag{11}$$

with $\sigma$ being the projected surface area averaged over many orientations. Fig. 2 shows a representative aggregate, with $R_\sigma$ and the outer radius $R_{max}$ indicated. The fluffiness of the aggregate is $1 - \Phi_\sigma$; in other words, a lower compactness factor correlates to a fluffier structure.

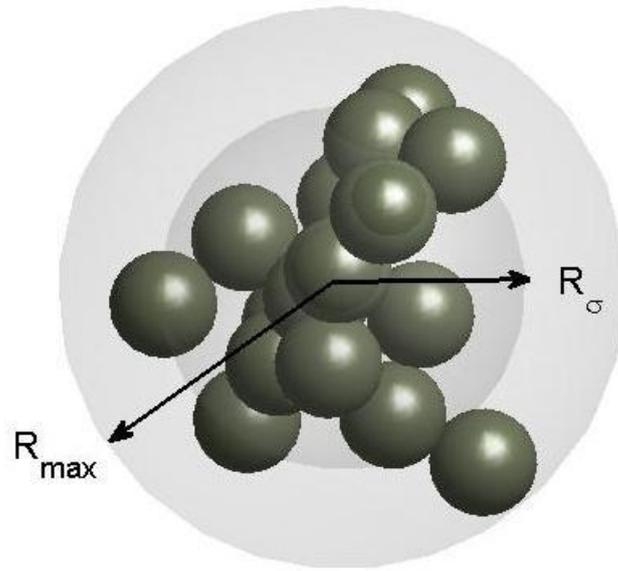

Fig. 2. Illustration of the compactness factor. The inner shaded area corresponds to a sphere with radius $R_\sigma$, the outer shaded area to a sphere with maximum radius, $R_{max}$. For compact aggregates, the volumes of the two spheres are approximately equal. For open aggregates, the ratio of the two approaches zero.

## 3. RESULTS

### 3.1. Charging of Aggregates

Before estimating the equilibrium surface charge on the aggregates, the time to reach the equilibrium condition needs to be considered, for depending on the plasma parameters and the dynamic processes being considered, the equilibrium condition is not always satisfied for grains of all sizes.

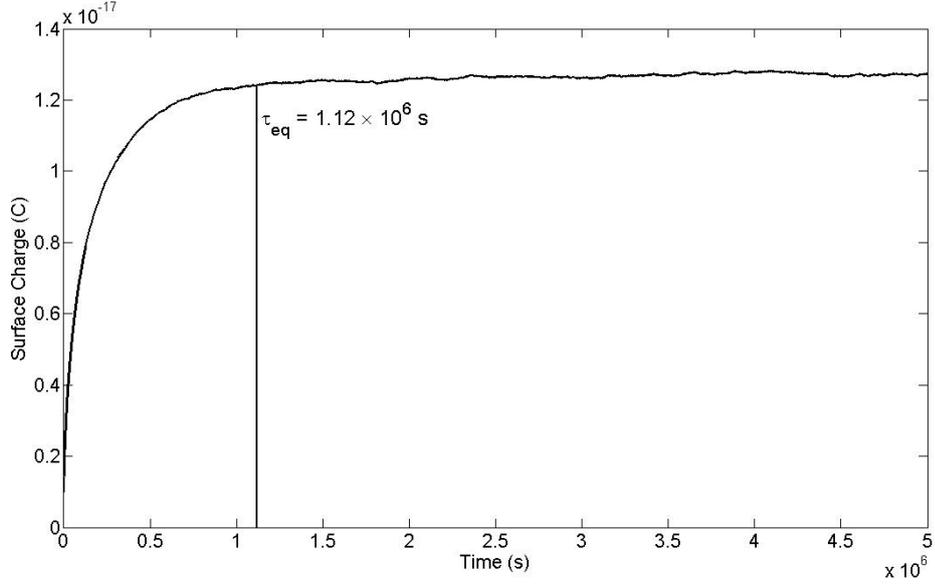

Fig. 3. Charging curve for a dimer consisting of two 5 nm-radius monomers. $\tau_{eq}$ is determined by the point where the absolute change in the charge is less than 0.1% of the equilibrium charge.

The dominant current determines the polarity of the equilibrium charge, while the non-dominant current determines the charging time, $\tau_{eq}$. As the grain charges, the relative contribution of the non-dominant current increases to balance the dominant current. Due to the high temperature of the plasma near the heliopause, secondary electron emission is the dominant charging process, determining $Q$. Thus $\tau_{eq}$ can be approximated by $|Q/J_e|$. Generally, $\tau_{eq}$ increases with decreasing dust radii $a$, approximately according to $\tau_{eq} \propto a^{-1}$. Fig. 3 shows the charging history of a dimer consisting of two 5 nm-radius monomers, the smallest aggregate in the simulation. The maximum charging time is approximately $1 \times 10^6$ s, which is less than $5.75 \times 10^6$ s, the time needed for interstellar dust grains to travel 1 AU with a constant speed of 26 km s$^{-1}$ (Schwenn 1990). Thus, all the aggregates in the simulation are assumed to reach equilibrium within traveling a distance of 1 AU.

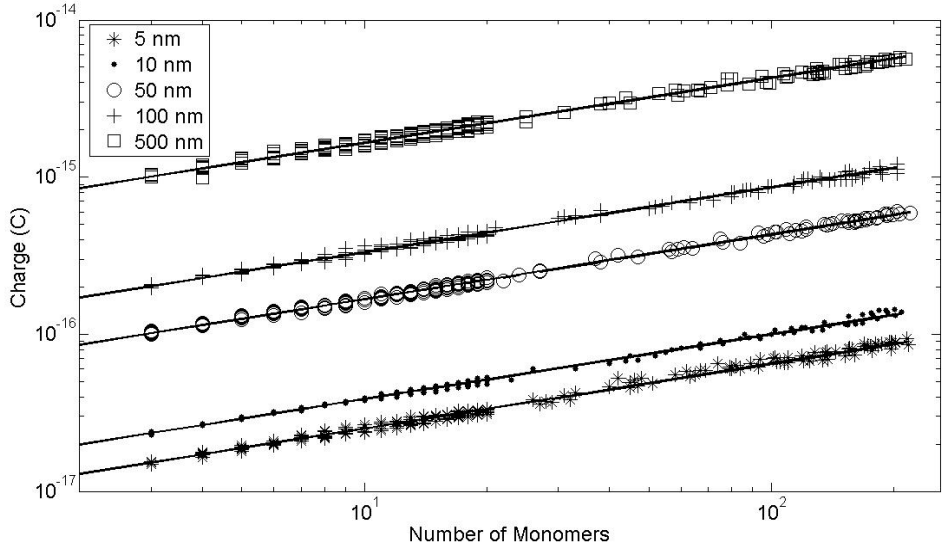

Fig. 4. Surface charge on aggregates as a function of the number of constituent monomers. The linear fits have the same slope for all monomer sizes with standard error less than 2%.

The calculated equilibrium surface charge on aggregates as a function of the number of monomers is shown in Fig. 4. The aggregates in each group consist of mono-disperse monomers up to $N = 200$. The legend indicates the sizes of the constituent monomers. The charge on each group of aggregates can be approximated as

$$Q_{agg} = Q_0 N^{0.413}, \qquad (12)$$

where $Q_0$ is the charge an isolated spherical monomer acquires, and $N$ is the number of monomers within an aggregate. The fact that the total charge on the aggregate is not simply the sum of the charge on each individual constituent monomer demonstrates that the aggregate's fluffy structure has significant influence on the overall charging process. Although a clear trend is evident indication that the aggregate charge increases with the number of monomers, it is difficult, if not impossible, to determine the number of monomers within an aggregate measured *in*

*situ*. Thus, a more fundamental parameter which characterizes the structure of the aggregate and is more easily determined needs to be determined. This is discussed in Section 3.2.

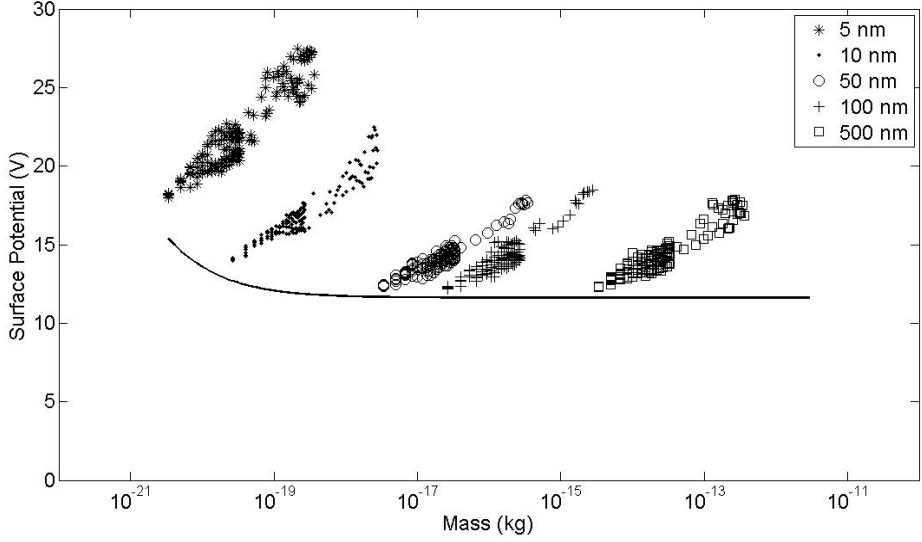

Fig. 5. Comparison of the surface potential on aggregates (data points) and spheres having the same mass (solid line).

In a given plasma environment, spherical grains with $a > 10$ nm reach the same equilibrium potential, independent of their radii. On the other hand, the small particle effect from secondary electron emission will cause spheres with radius $a < 10$ nm to have a greater potential (Chow et al. 1993). However, as shown in Fig. 5, the surface potential of an aggregate clearly does not follow this trend. The surface potential of an aggregate in this study is defined as

$$\varphi = \frac{Q_{agg}}{4\pi\varepsilon_0 r_{mass}}, \qquad (13)$$

where $Q_{agg}$ is the total charge on the aggregate, and $r_{mass}$ is the radius of a solid silicate sphere having the same mass as the aggregate. Overall, the surface potential of aggregates shows greater fluctuation and is generally higher than that of the spheres, due to the greater surface area of the aggregate compared to a sphere having the same mass. The similar effect has also been shown in a recent experimental study (Wiese et al. 2010). Meanwhile, aggregates consisting of monomers with *a* = 5 nm and 10 nm have a surface potential which is significantly higher than that for a sphere with the same mass. This is caused by the high positive charge each constituent monomer carries as a result of the small particle effect, and is consistent with Kimura & Mann's prediction (1998).

**3.2. Characterization of Aggregates Using Compactness Factor**

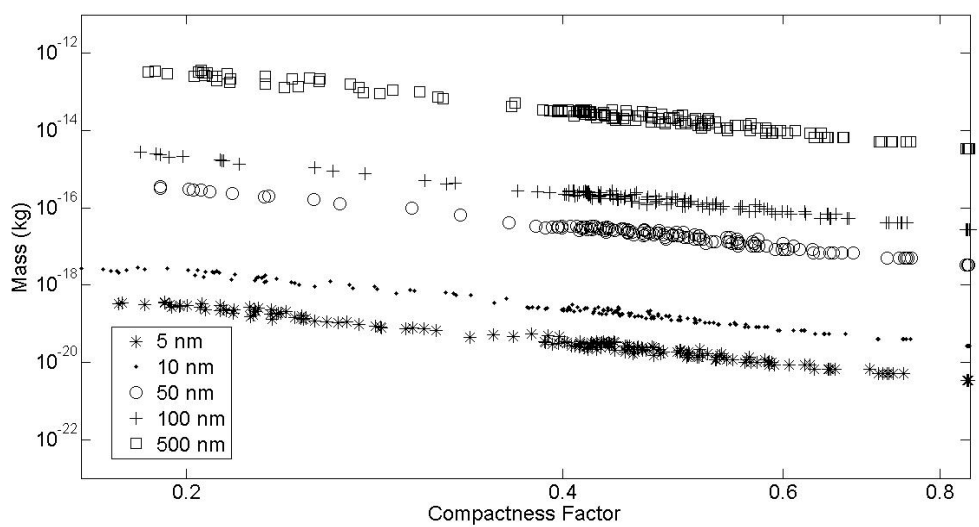

Fig. 6. Mass of aggregates as a function of the compactness factor.

Since the charge on an aggregate is determined by the surface area of the aggregate, which is related to the fluffiness of the aggregate structure, a fundamental parameter characterizing the aggregate structure would serve as a better candidate for predicting the aggregate charge. The mass of aggregates ranging in size from two to 200 monomers is shown as a function of the compactness factor $\Phi_\sigma$ in Fig. 6. For each of the groups, as the mass increases, the compactness factor decreases, indicating a fluffier structure. The linear relationship on a log-log plot implies that the compactness factor well characterizes the structure of the aggregates. The equilibrium surface charge as a function of the compactness factor is shown in Fig. 7 (a) for aggregates. Each group can be fit with a straight line of the same slope on a log-log plot with surface charge being given by

$$Q_{agg} = \frac{Q_0}{\Phi_\sigma^{1.3}} = \frac{4\pi\varepsilon_0 a\varphi}{\Phi_\sigma^{1.3}}, \qquad (14)$$

where $a$ is the radius of a constituent monomer, and $\varphi$ is the surface potential of the monomer of radius $a$. The results clearly demonstrate that for each constituent size, the surface charge on the aggregate increases as the fluffiness of the aggregate structure increases, with the surface charge of a sphere being the lower limit as $\Phi_\sigma$ approaches one. Using the surface potential of a sphere of an equivalent mass to calculate the charge on an aggregate thus leads to an underestimate of the charge, as indicated in Fig. 5.

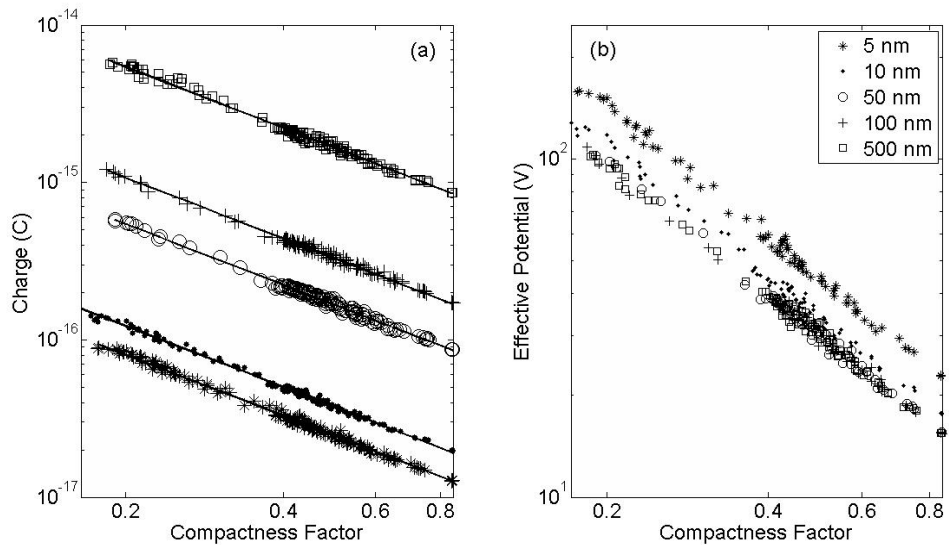

Fig. 7. (a) The surface charge on aggregates consisting of mono-disperse monomers of different radii,

and (b) the surface charge divided by the capacitance of a single monomer. The small particle effect is clearly evident for aggregates composed of monomers of size $a$ = 5 nm.

In Fig. 7(b), the charges on the aggregates are divided by $4\pi\varepsilon_0 a$, the capacitance of a single monomer, to yield the effective potential for the aggregates. After eliminating the size factor, it is clear that aggregates consisting of monomers with $a$ > 10 nm fall on the same line, while the ones with the smallest monomers, $a$ = 5 nm, exhibit substantially higher y-intercept. The aggregates consisting of monomers with $a$ = 10 nm lie between, clearly indicating that the small particle effect is most pronounced for particles with radius $a$ < 10 nm.

### 3.3. Charge-to-mass Ratio

Charged interstellar dust grains will be deflected under the influence of the Lorentz force when flowing from the heliopause to the termination shock. The magnitude of this deflection depends on the dust grain velocity, the magnetic field, and the grain's charge-to-mass ratio. The high positive charge that small dust grains obtain near the heliopause can lead to large scattering, with a sufficiently large deflection leading to the depletion of these small grains within the heliosphere. An estimate of the influence of the Lorentz force on dust dynamics can be found using the gyroradius

$$r_g = \frac{mv}{QB}, \qquad (15)$$

where $Q$ and $m$ are the charge and mass of the grain, and $v$ is the speed of the grain, taken to be 26 km s$^{-1}$. The magnetic field near the heliopause is $B$ = 1.3 nT, transverse to the flux of interstellar dust (Nerney et al. 1993).

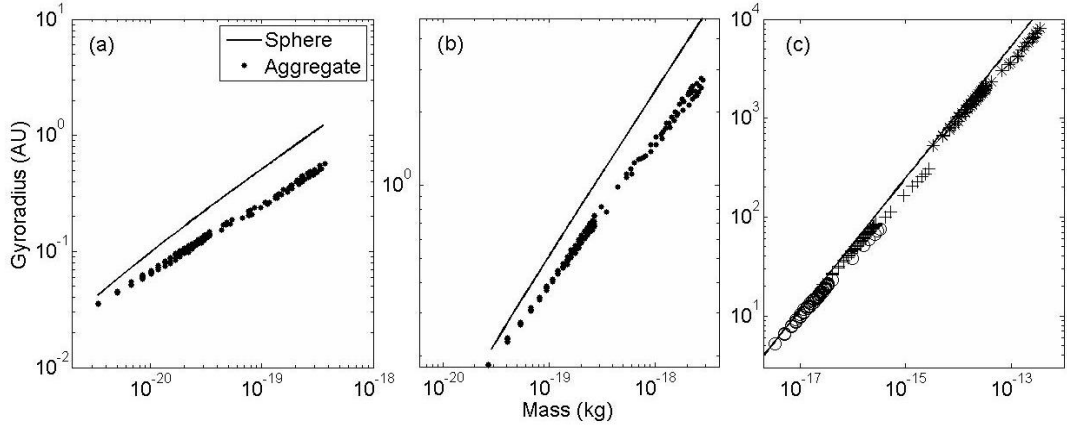

Fig. 8. The gyroradii computed for aggregates consisting of monomers with (a) radius $a = 5$ nm, (b) $a = 10$ nm, and (c) $a = 50$ nm (sphere), 100 nm (cross) and 500 nm (star). The gyroradii for the spheres having the same mass is plotted for comparison (straight line).

The gyroradii of aggregates and single spheres are plotted as a function of mass in Fig. 8. For a single sphere, $r_g \propto m^{2/3}$. The gyroradii for aggregates differ from that found for spheres since the charge on the aggregates is highly dependent on the overall aggregate structure. In Fig. 6 it is shown that aggregates with greater mass have a fluffier structure, and thus are more highly charged. This is also evident as shown in Fig. 8; as the mass increases, the gyroradius deviates from that found for a sphere of the same mass, implying a higher charge-to-mass ratio. Fig. 8 (a) and (b) show the gyroradii for aggregates consisting of monomers with $a = 5$ nm and 10 nm, respectively. The large discrepancy between the gyroradii of the aggregates and those of the spheres is due to the enhanced small particle effect. For aggregates consisting of monomers with $a > 10$ nm, the gyroradii show less difference from that found for spheres. These results indicate that the gyroradius of a sphere may be considered the upper limit for aggregates, with aggregates having larger charge-to-mass ratios.

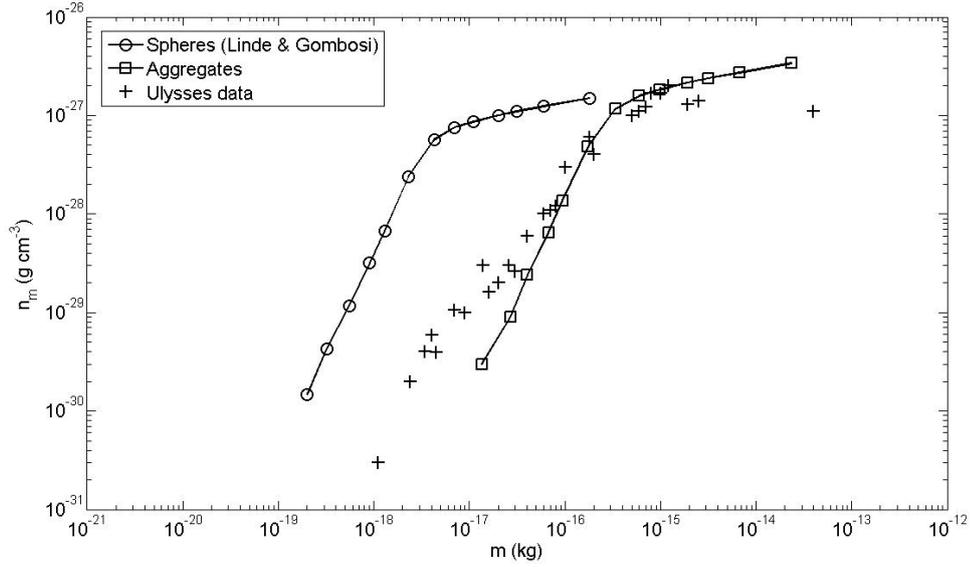

Fig. 9. Comparison of the filtration of MRN distribution for interstellar dust near the heliopause given by aggregates (square), spheres (sphere, Linde & Gombosi), and Ulysses data (cross). The aggregates used for the present model consist of 50 nm-radius monomers.

It is interesting to examine how the aggregate's charge-to-mass ratio affects interstellar dust filtration rate around the heliopause. Although the quantitative results can only be obtained through detailed simulations, a crude estimate can be obtained by applying a filtration rate to the Mathis, Rumpl and Nordsieck (MRN) distribution for interstellar dust (1977) and comparing it to Ulysses data. Linde & Gombosi have shown that the cutoff charge-to-mass ratio lies at around 0.3-0.5 C kg$^{-1}$, and derived the filtered MRN distribution in the heliosphere assuming they are homogenous spheres (Linde & Gombosi 2000). This translates to a cutoff mass of $2 \times 10^{-18}$ kg as shown in Fig. 9. As a first order approximation, we consider the charge-to-mass ratios of aggregates consisting of 50 nm-radius monomers. The aggregate charge can be obtained through Eq. (12), and the corresponding mass through

$$m_{agg} = Nm_0, \qquad (16)$$

where $N$ is the number of monomers, and $m_0$ is the monomer mass. The filtration rate is given by Linde & Gombosi (Fig. 7, 2000), and is applied to MRN distribution. The higher charge-to-mass ratio of the aggregates shifts the cutoff mass to higher mass, and yields a better fit for Ulysses data, although it fails to fit both the small mass particles and large mass particles detected by Ulysses.

Three explanations are possible for this failure. First, mono-dispersed monomers are used in the current simulation. However, interstellar dust grains are generally expected to have a poly-disperse distribution (Mathis et al. 1977), and it is not unreasonable to assume that aggregates consisting of poly-dispersed monomers would have a different charge-to-mass ratio than aggregates with mono-dispersed monomers. Second, silicate is used as the constituent of interstellar dust. Based on remote observations and simulation, interstellar grains are more likely to have a silicate-core, organic-coated structure (Greenberg & Hage 1990; Kimura et al. 2003). This would change the electrical properties of interstellar dust, thus altering the charge-to-mass ratio. Third, the upper limit for MRN distribution is uncertain since large grains make only minor contributions to the interstellar extinction curve (Mathis et al. 1977). Thus, an alternative distribution rather than MRN may be needed to account for particles with $m > 1 \times 10^{-15}$ kg.

## 4. DISCUSSION AND CONCLUSION

In this study, the charge on aggregates near the heliopause is calculated including plasma current and secondary electron emission. It is shown in Fig. 7 and 8 that the small particle effect, in which the grain charge is enhanced for monomer radii $a < 10$ nm, becomes significant for aggregates consisting of many nano-sized particles, and that aggregates generally have a higher charge-to-mass ratio compared to spherical grains. With the introduction of a compactness factor, the charge on aggregates can be related to its structural characteristics, as shown in Fig. 7 (a). This provides a new approach to estimating the charge-to-mass ratio of interstellar dust grains through remote observation since the optical properties of interstellar grains also depend on their morphology (geometry, refractive index, etc). In fact, Shen et al. (2008) have recently shown a dependence of the extinction cross section on the porosity of the aggregates. As such, the compactness factor may serve as a useful tool when investigating the dynamics of interstellar dust grains in the outer heliosphere.

Using the aggregate's charge-to-mass ratio yields a better fit for Ulysses data (Fig. 9), which serves as strong evidence that the aggregate structure of interstellar grains, with their higher charge-to-mass ratios, is important for the dynamics of these grains. Although the composition, structure and size of interstellar dust are not fully understood, the current study presents a preliminary model to estimate the charge on the aggregate, and serves as a useful tool when simulating the dynamics of interstellar grains. Future work will include simulation and analysis of interstellar dust charge assuming a poly-disperse distribution in order to obtain a more accurate picture of the charging of interstellar dust grains and its consequences for grains' coagulation and dynamics.

**ACKNOWLEDGEMENT:**

This material is based upon work supported by the National Science Foundation under Grant No. 0847127.

**REFERENCES:**

Blum, J., & Wurm, G. 2000, Icarus, **143**, 138
Chow, V. W., Mendis, D. A., & Rosenberg, M. 1993, J. Geophys. Res., **98**, 19065
Désert, F. X., Boulanger, F., & Puget, J. L. 1990, A&A, **237**, 215
Draine, B. T., & Lee, H. M. 1984, ApJ, **285**, 89
Draine, B. T., & Salpeter, E. E. 1979, ApJ, **231**, 77
Goertz, C. K.. 1989, Rev. Geophys., **27**, 271
Greenberg, J. M., & Hage, J. I. 1990, ApJ, **361**, 260
Grün, E., et al. 1993, Nature. **362**, 428
Kimura, H., & Mann, I. 1998, ApJ, **499**, 454
Kimura, H., & Mann, I. 2000, Adv. Space Res., **25**, 299
Kimura, H., Mann, I., & Jessberger, E. K. 2003, ApJ, **583**, 314
Krüger, H., et al. 2001, Planet Space Sci. **49**, 1303
Laframboise, J. G., & Parker, L. W. 1973, Phys. Fluids., **16**, 629
Levy, E. H., & Jokipii, J. R. 1976, Nature, **264**, 423
Linde, T. J., & Gombosi, T. I. 2000, J. Geophys. Res., **105**, 10411
Mathis, J. S., & Whiffen, G. 1989, ApJ, **341**, 808
Mathis, J. S., Rumpl, W., & Nordsieck, K. H. 1977, ApJ, **217**, 425
Matthews, L. S., et al. 2007, IEEE Trans. Plasma Sci., **35**, 260
Matthews, L. S., & Hyde, T. W. 2008, IEEE Trans. Plasma Sci., **36**, 310
Matthews, L. S., & Hyde, T. W. 2009, New J. Phys. **11**, 063030
McCarthy, J. F., et al. 1980, ApJ, **242**, 965
Mukai, T. 1981, A&A, **99**, 1
Nerney, S., Suess, S. T., & Schmahl, E. J. 1993, J. Geophys. Res., **98**, 15169
Paszun, D., & Dominik, C. 2009, A&A, **507**, 1023
Pauls, H. L., & Zank, G. P. 1996, J. Geophys. Res., **101**, 17081
Savage, B. D., & Mathis, J. S.,1979, Ann. Rev. Astron. Astrophys., **17**, 73
Schwenn, R. 1990, in Physics and Chemistry in Space 20, Physics of the Inner Heliosphere I: Large-Scale Phenomena, ed. R. Schwenn & E. Marsch (Berlin: Springer), 99
Shen, Y., Draine, B. T., & Johnson, E. T. 2008, ApJ, **689**, 260
Wiese, R., et al. 2010, New J. Phys. **12**, 033036
Whipple, E. C. 1981, Rep. Prog. Phys., **44**, 1197
Wurm, G., & Blum, J. 1998, Icarus, **132**, 125
Woo, J. W., et al. 1994, ApJ, **436**, L5